\documentclass[a4paper]{article}

\usepackage{icrc2013}
\usepackage[english]{babel}

\title{Recent Highlights from IceCube}

\shorttitle{Highlights from IceCube}

\authors{
Spencer Klein$^{1,2}$
for the IceCube Collaboration$^{3}$.
}

\afiliations{
$^1$ Lawrence Berkeley National Laboratory, Berkeley, CA, 94720 USA \\
$^2$ University of California, Berkeley, CA, 94720 USA \\
$^3$ The full collaboration author list is available at http://icecube.wisc.edu/collaboration/authors/current
}

\email{srklein@lbl.gov}

\abstract{The $\sim$1 km$^3$ IceCube neutrino observatory was completed in December, 2010 and is taking data on cosmic-ray muons and neutrinos, extra-terrestrial neutrinos, and setting limits on a variety of exotic phenomena.   This proceeding will cover recent IceCube results, with an emphasis on cosmic-rays and on searches for extra-terrestrial neutrinos, with a stress on results that were presented at the 2013 International Cosmic Ray Conference. 
}


\begin{document}
\maketitle

\section{Introduction}

The IceCube neutrino observatory collects data on a wide variety of topics involving cosmic-rays and astrophysics: cosmic-ray air showers, including their high-energy muon content, atmospheric and astrophysical neutrinos, and a variety of searches for exotica, including neutrinos from dark matter annihilation, magnetic monopoles, and signs of supersymmetric particle production in neutrino interactions.   This writeup presents highlights from recent IceCube activities in these areas, with some focus on results that were presented at the 2013 International Cosmic Ray Conference.   This work was performed by the $\sim$250 member IceCube Collaboration, including scientists and engineers from the around the world.  I will present data taken with a variety of different IceCube configuration, denoted by ICxx, where xx is the number of active strings in each configuration.  Table 1 of Ref. \cite{Resconi:2013ola} discusses some of  the characteristics of the different configurations. 

\section{IceCube: hardware and performance}

The IceCube neutrino observatory is a $\sim$1 km$^3$ detector that observes the Cherenkov radiation from electrically charged particles, including those produced in neutrino interactions \cite{Halzen:2010yj}.  It is deployed at the South Pole, about 1 km from the Amundsen-Scott South Pole Station.  The Cherenkov radiation is detected with 5,160 optical sensors (digital optical modules, or DOMs) which are mounted on 86 vertical cables (strings) emplaced in the ice.  On  78 of the strings, DOMs are mounted every 17 m between 1450 and 2450 m. These strings are on a 125 m triangular grid.  In the DeepCore subarray, the string and DOM spacings are considerably smaller \cite{Collaboration:2011ym}.  The main array has an energy threshold around 100 GeV, while DeepCore observes neutrinos with energies as low as 10 GeV.

In addition to the buried DOMs, IceCube incorporates a surface array called IceTop which consists of 162 ice filled tanks that detect the Cherenkov radiation from charged particles in cosmic-ray air showers  \cite{IceCube:2012nn}.   The tanks are deployed in pairs (called `stations'), with most stations deployed near the top of the IceCube strings.  The tanks are 182 cm in diameter, filled to a depth of 90 cm with ice.  Two DOMs in each tank observe the Cherenkov light from charged particles in air showers. IceTop is sensitive to showers with energies above 100 TeV. The 1 km$^2$ surface area is large enough to observe showers with energies up to 1 EeV.  The top of the South Pole icecap is 2835 m above sea level, so IceTops elevation is relatively close to the predicted shower maximum for PeV showers. 

Each DOM consists of a 25.4 cm(10 inch) Hamamatsu R7081 photomultiplier tube (PMT) \cite{Abbasi:2010vc}, plus data acquisition electronics in a spherical glass pressure housing.  DOMs operate autonomously, sending packetized digital data to the surface \cite{Abbasi:2008aa}, and are largely self-calibrating.   The data acquisition system must accurately record the arrival times of most photoelectrons while running reliably at very low temperatures with very low power consumption.  Each DOM includes a precision crystal oscillator which is calibrated by a system which exchanges timing signals between the surface and the DOMs.  It maintains timing calibrations of about 2 nsec, across the entire array \cite{Achterberg:2006md}.  The hardware has been extremely reliable.  Currently, 98.5\% of the DOMs are taking data.  Of the failures, the vast majority were `infant mortality' during deployment. Only 2 DOMs failed during 2012.  The detector has also run very reliably, with a typical up-time over 99\%. 

The data acquisition system includes two waveform digitizer subsystems.  The first comprises a custom switched capacitor array (SCA) ADC, running at 300 MS/s and taking 128 samples (400 ns) per trigger.  Each phototube feeds three SCA channels, with varying gains, providing 14 bits of dynamic range.  The second subsystem comprises a 10-bit commercial ADC, running at 40 MS/s, and collecting 6.4 $\mu$s of data for each trigger.

Considerable effort has gone into understanding the optical behavior of the active medium - the ice.  The Antarctic ice is extremely pure, with typical effective scattering lengths around 20-25 m, and absorption lengths around 100 m; in the clearest ice, the absorption length is over 200 m.  This purity allows DOMs up to several hundred meters from an interaction to collect useful data, but it also challenges the calibration methods.  The absorption and scattering lengths depend on  the wavelength of the light, and the position in the ice.   The wavelength dependence was studied in IceCubes predecessor AMANDA \cite{AMANDA}, using LEDs and lasers that emitted at different frequencies \cite{AMANDAice}.  The spatial variation of the optical properties is studied in IceCube using an ensemble of LEDs (each DOM contains 13 LEDs) and two 337 nm N$_2$ lasers, plus data from downward-going atmospheric muons \cite{Aartsen:2013rt}.  We have recently found that the scattering and absorption lengths are anisotropic \cite{Dima}, possibly because dust grains in the ice align along the direction of ice flow. 

\begin{figure}[!t]
  \centering
\includegraphics[width=3 in]{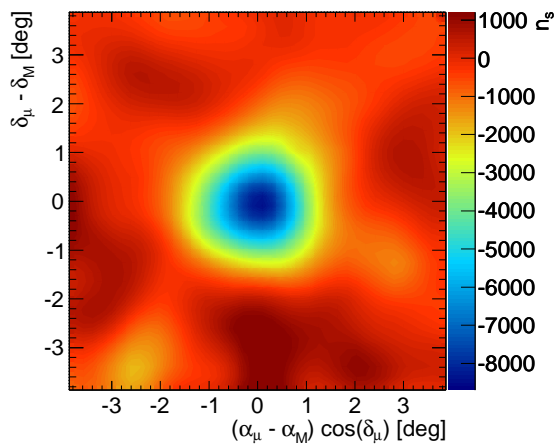}
  \caption{Contour plot showing the deficit (in number of events) as a function of the displacement from the center of the Moon, using IC59 data.}
  \label{fig:moonshadow}
 \end{figure}

IceCube also uses cosmic-ray muons for higher level calibrations.  One study takes advantage of the fact that the Moon blocks cosmic-rays, creating a `hole' in the sky which is visible in the downward-going muon rate.  Figure \ref{fig:moonshadow} shows the shadow of the moon in the IC59 data; IC40 shows a similar deficit.  The width and depth of the deficit match IceCube simulations.  The center of the hole is within 0.2$^0$ of the actual position of the Moon \cite{Aartsen:2013zka}, confirming IceCubes pointing accuracy.

Each of the three neutrino flavors, $\nu_e$, $\nu_\mu$ and $\nu_\tau$, has a specific signature in IceCube.  Charged current $\nu_\mu$ produce a hadronic shower, typically containing 20\% of the neutrino energy, with the rest going into an energetic muon which leaves a long track in IceCube.  These events have very good angular resolution, but, because muons can carry energy into or out of the detector, limited energy resolution, typically a factor of two in energy.  Energetic (above 1 PeV) charged current $\nu_\tau$ produce a characteristic 'double-bang' signature - one shower when the $\nu_\tau$ interacts, and another, 100 m or so away, when the $\tau$ decays \cite{Learned:1994wg}.  At somewhat lower energies, $\nu_\tau$ may produce a `double-pulse' signature, where the two showers produce hits in the same DOMs, but at different times 
\cite{Cowen:2007ny}.  Charged current $\nu_e$, lower-energy charged current $\nu_\tau$ and all flavors of neutral current interactions produce electromagnetic or hadronic showers which, except at energies above 10 PeV, are nearly point sources of light.  IceCube can accurately measure the energy of these events, but the angular determination is much poorer.

\section{Cosmic-rays}

\begin{figure}[!t]
  \centering
\includegraphics[width=3 in]{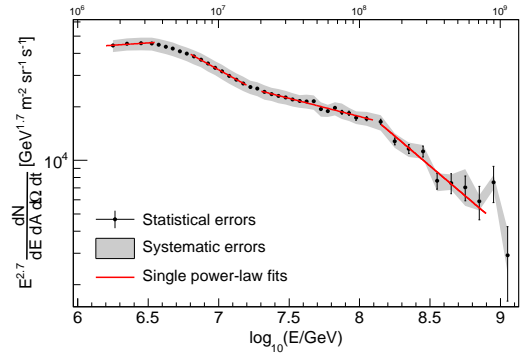}
  \caption{The cosmic-ray energy spectrum measured by IceTop-73.  In addition to the knee around 4 PeV, additional spectral breaks are seen around 18 and 130 PeV.}
  \label{fig:CRspectrum}
 \end{figure}

IceCube has measured the cosmic-ray energy spectrum at energies between 1.58 PeV and 1.26 EeV using data from 73 IceTop stations \cite{CRspectrum}.  The analysis used events where at least 5 stations recorded a hit.  As Fig. \ref{fig:CRspectrum} shows, in this energy range, we observe three spectral breaks.  In addition to the knee, at 4 PeV,  the spectrum hardens around 18 PeV, and then softens around 130 PeV. The analysis considered 5 possible cosmic-ray compositions, including pure protons, pure iron, and three mixed compositions.    The composition sensitivity was studied and constrained by the requirement that the measured cosmic-ray flux be independent of the zenith angle.  The analysis used a mixed composition \cite{Gaisser:2012zz} which fulfilled that criteria reasonably well.  The degree of flux mismatch at different zenith angles was used to estimate systematic errors.   There is also some systematic uncertainty due to the varying depths of snow atop the IceTop tanks\cite{CRice}.  Future work will use events with fewer hit stations, pushing the spectral measurement down to 100 TeV. 

\begin{figure}[!t]
  \centering
\includegraphics[width=3 in]{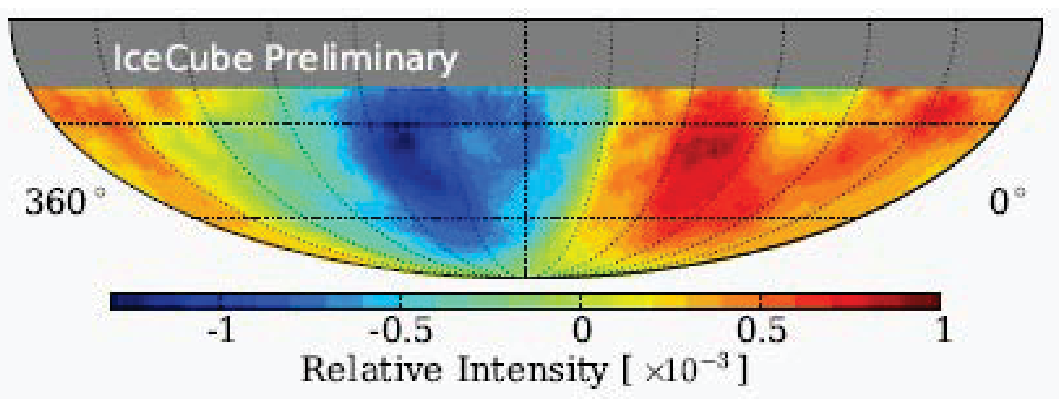}
\includegraphics[width=3 in]{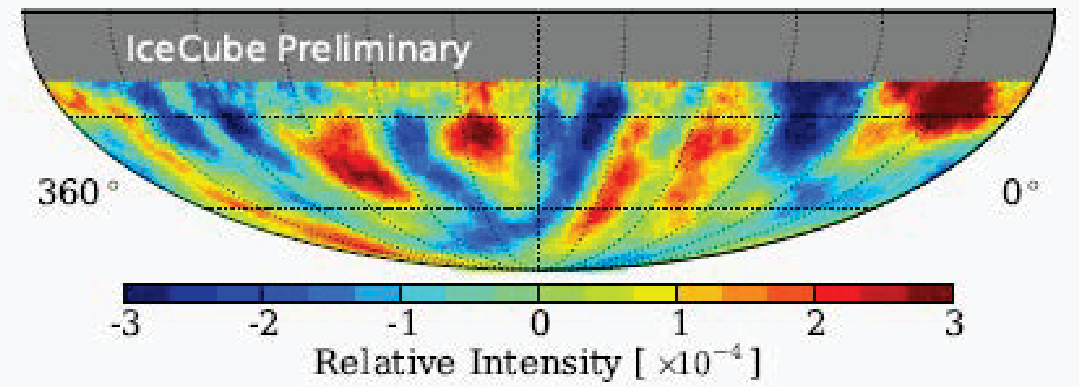}
  \caption{The cosmic-ray anisotropy observed using 5 years of data, containing 150 billion events.  The top panel shows the intensity of the raw anisotropy, with the bottom showing the result after the dipole and quadrupole moments are subtracted \cite{CRanisotropy}. }
 \label{fig:CRanisotropy}
 \end{figure}

One surprising aspect of the cosmic-ray flux is that it is slightly anistropic.  IceCube has measured the anisotropy in the Southern hemisphere, complementing Northern sky studies by the Tibet air shower array, Super-Kamiokande, and Milagro.  Separate studies were done with the IceTop surface array and using muons observed in the in-ice detector \cite{CRanisotropy}.  The muon channel, shown in Fig. \ref{fig:CRanisotropy}, offers much higher statistics (150 billion events), covering the energy range from 20 to 400 TeV.
The IceTop studies had lower statistics, but covered a higher energy range; data was analyzed in two bins, with median energies of 400 TeV and 2 PeV.  The typical intensity fluctuations were a few parts in 1,000, and they were present over a wide range of angular scales.  The anisotropy distribution changes with energy,  but the fractional strength of the fluctuations are comparable, within a factor of 2.   This challenges current models of cosmic-ray production and propagation; one possibility is that some of the variation is due to one or more relatively local sources. 

IceCube has also measured the cosmic-ray composition at energies between 1 PeV and 1 EeV using events with showers in the IceTop array and muons in the in-ice detector.  For a given energy, heavier primary particles produce more muons.  The in-ice detector has a threshold for single muons of  around 500 GeV; these muons are produced much earlier in the shower than the lower energy muons observed by surface detectors.   IceCube measures the light produced by the muon bundle near the shower core; this light is proportional to the total energy of the muons in the bundle, which is statistically related to the shower composition using simulations.  The average atomic number, $\langle A \rangle$, rises with increasing primary energy, up to an energy of about 100 PeV \cite{IceCube:2012vv}.  

\begin{figure}[!t]
 \centering
\includegraphics[width=3 in]{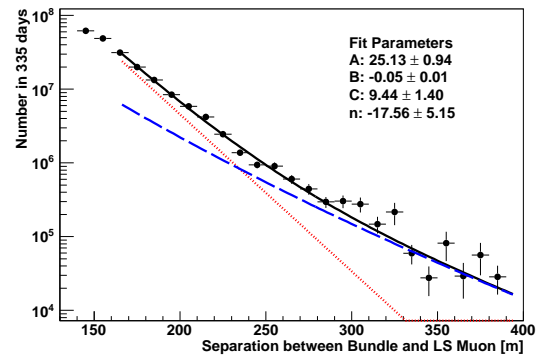}
\caption{Lateral separation spectrum of high transverse momentum muons observed in IceCube.  The data points are fit to an exponential, expected for muon production in soft hadronic interactions, plus a power law, as expected at large separations in perturbative QCD \cite{Abbasi:2012kza}.}
\label{fig:lateralmuon}
\end{figure}

IceCube has also studied isolated muons in cosmic-ray air showers at distances between 135 m and 400 m from the shower core \cite{Abbasi:2012kza}.  This measurement extends the MACRO lateral separation spectrum  \cite{MACRO} by an order of magnitude in distance.  At distances over 100 m, most of the lateral separation comes from the initial transverse momentum, $p_T$, of the muon with respect to the initial cosmic-ray direction.  The $p_T$ is
\begin{equation}
p_T = \frac{dE_\mu}{h}.
\end{equation}
Here $d$ is the lateral separation of the muon from the shower center, $h$ is the
distance from the primary interaction (typically 30 km for vertically incident showers), and $E_\mu$ is the muon energy.   Fig \ref{fig:lateralmuon} shows the measured lateral separation spectrum.   The minimum visible IceCube separation of 135 m corresponds roughly to $p_T > 2$ GeV/c, the regime where perturbative quantum chromodynamics (pQCD) calculations are used at colliders like the Large Hadron Collider or RHIC.  The data exhibits a transition from an exponential decrease, as expected for models of soft (non-pertubative) hadron production, to a power law, as predicted by pQCD.  This opens the door to pQCD-based studies of phenomena like the cosmic-ray composition. 

IceCube also studies cosmic-rays from closer sources.  On May 17, 2012, an increase in the particle rates in all IceTop tanks was observed.   This increase coincided with a solar flare which was observed by other ground-level neutron monitors, and the GOES-13 satellite \cite{Solarflare}.  A similar increase was observed on December 13, 2006 \cite{Abbasi:2008vr}.
The enhancement is consistent with expectations from low-energy cosmic-rays coming from the Sun.  Fits to the detector rate increase vs. tank threshold indicate that most of the cosmic-rays have energies below 1 GeV.  

\section{WIMPs and other physics}

In addition to neutrinos and cosmic rays, IceCube studies many other physics topics, including searches for particle dark matter, bursts of $\sim$10 MeV neutrinos from supernovae, magnetic monopoles, and searches for signatures of new physics, such as pairs of upward-going particles.

In many models, dark matter particles are their own antiparticles, so are self-annihilating.  They may be captured in gravitational wells. There their density increases over time until the rate of self-annihilation equals the capture rate, and equilibrium is reached.  The products of the annihilation depend on the details of the model.  IceCube considers the following channels: $W^+W^-$ ($\tau^+\tau^-$ below threshold), $b\overline b$, and sometimes $\mu^+\mu^-$ and $\nu\overline\nu$.  The $\nu\overline\nu$ produces the hardest neutrino spectrum and $b\overline b$ the softest.  We have searched for neutrinos from WIMP annihilation coming from the Sun \cite{Aartsen:2012kia}, the galactic center and halo, and nearby dwarf galaxies.   

The Sun is a particularly interesting target since it is composed mostly of hydrogen.  This is the ideal material to attract WIMPs with spin-dependent couplings to matter.  WIMPs passing through the Sun scatter and are gravitationally captured by it.  Their density rises and they begin colliding and annihilating.  Based on our non-observation of neutrinos coming from the Sun, we have set limits on various models for weakly interacting dark matter \cite{Aartsen:2012kia}.  These limits constrain many variants of supersymmetry that predict spin-dependent WIMP-matter couplings.  We are also searching for neutrinos from WIMP annihilation in the Earth \cite{EarthWIMP}.  Here, the capture cross-section increases when the WIMP mass is comparable in mass to the nuclei that comprise the Earth.  For example, there is an expected peak around 50 GeV, corresponding to iron nuclei.  

\begin{figure}[!t]
  \centering
\includegraphics[width=3 in]{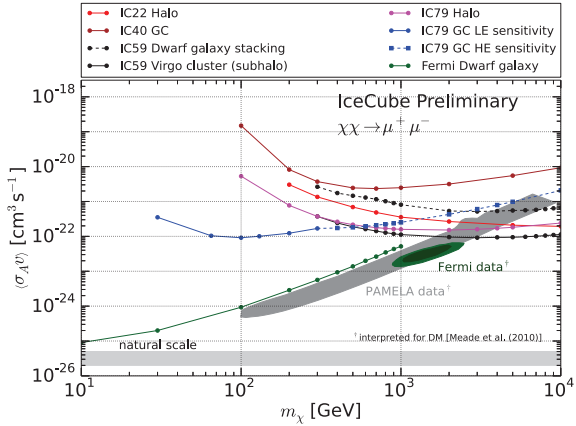}
 \caption{Limits on the product of the WIMP annihilation cross-section and mean velocity, $\langle \rho v\rangle$, for different IceCube analyses, for annihilation into the final state $\mu^+\mu^-$.  Limits are shown for annihilation in the galactic halo (``Halo"), galactic center (``GC"; limits from three different searches are shown), nearby dwarf galaxies and the Virgo cluster \cite{Abbasi:2011eq}. Also shown is the limit from the Fermi observatory \cite{Fermi}, along with the expectations if the Pamela excess is interpreted as being due to dark matter \cite{DM}. The lower, light grey band shows the naturalness scale, at which WIMPs may be present as thermal relics. }
 \label{fig:WIMP}
 \end{figure}

For other targets, the WIMPs annihilate `in-flight.'  We set limits on the product of the dark matter particle density and mean velocity $\langle \rho v\rangle$. Figure \ref{fig:WIMP} compares the IceCube limits on $\langle \rho v\rangle$, as a function of WIMP mass, with those obtained by the PAMELA and FERMI experiments \cite{WIMPpaper}.  The IceCube sensitivity increases with increasing particle mass, due to the increase in neutrino cross-section (in IceCube) and increasing muon range.

Although individual neutrinos with energies of order 10 MeV do not trigger IceCube, a short burst, such as that from a nearby supernova, could be observable as a collective increase in the single photoelectron rates in all phototubes.  Because the dark noise rates in the IceCube DOMs are low (286 Hz),  the 5,160 DOMs operating together become a sensitive detector for MeV neutrinos, sensitive to supernova in all of the Milky Way, with more than $5\sigma$ significance.  There is also some sensitivity to supernovae in the Magellanic Clouds.  Although IceCube cannot determine the supernova energy spectrum, the time evolution of the signal can be measured with msec precision.  We are in the process of implementing two new techniques to boost our sensitivity.  The first is an improved analysis that will remove hits from observed muons, reducing the fluctuations in the measured PMT rates, and the second is an improved data acquisition system \cite{Abbasi:2011ss}, which will save all of the individual hits in the event of a suspected supernova. 

IceCube has completed  two types of searches for magnetic monopoles.  The first are searches for relativistic monpoles, which are moving fast enough to emit Cherenkov radiation.  Since monopoles are electromagnetically similar to electrically charged particles with $q=67.5$ e, these tracks produce abundant light.  We have set flux limits  that are a small fraction of the Parker bound for monopoles with velocities larger than 0.76 c \cite{Abbasi:2012eda}.  We also searched for slower monopoles, with 
velocities approximately in the range $10^{-3}$ to $10^{-4}$ c \cite{slowmonopoles}; these monopoles will only be visible if they catalyze proton decay.  We set limits on the flux vs. catalysis cross-section.

We are also searching for events with unusual topologies which are not expected in the standard model.  One example of this is a pair of parallel upward-going particles.  These events are predicted in some models of supersymmetry.  A neutrino interaction 1,000 km below IceCube can produce a pair of heavy charged supersymmetric particles, which propagate upward, separating as they go \cite{Albuquerque:2009vk}.  The standard model backgrounds for these processes appear to be small \cite{vanderDrift:2013zga}.

\section{Atmospheric neutrinos, oscillations and PINGU}

\begin{figure}[!t]
\centering
\includegraphics[width=3 in]{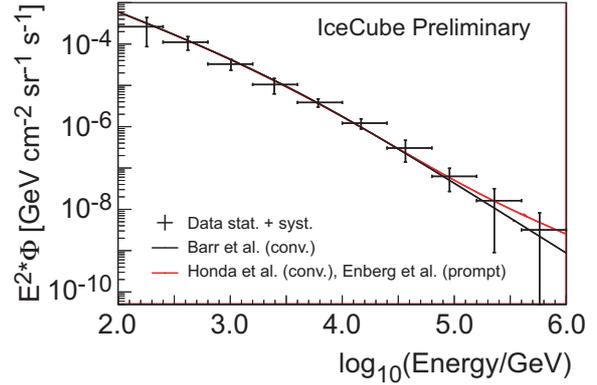}
\caption{The atmospheric neutrino spectrum measured using IC59, compared to two theoretical predictions, one with and one without a prompt component.  The prompt component is the central value of Ref. \cite{Enberg:2008te}.  }
\label{fig:IC59atmosphericnu}
 \end{figure}

IceCube observes a large number of atmospheric neutrinos each year.  Besides being a background to searches for extra-terrestrial neutrinos, they are of considerable physics interest in their own right.  

IceCube has measured the energy spectrum of $\nu_\mu$ from 100 GeV to 1 PeV \cite{atmosphericnu}, extending our previous measurement of the spectrum \cite{Abbasi:2010ie}.  The spectrum is determined based on measurements of muon specific energy loss ($dE/dx$) in the detector, with an unfolding procedure to account for the measurement resolution and the steeply falling spectrum.  Figure \ref{fig:IC59atmosphericnu} compares the measured spectrum with two theoretical models, with and without a prompt component.   Neutrino events were selected using a Random Forest machine learning approach.   It selected 27,771 neutrino events in 346 days of data with IC59; the purity of the sample is estimated to be 99.6\%.   
Multiple methods for the unfolding (and the required regularization) were tried; they all gave consistent results.   The error bars are dominated by the systematics, which were estimated with the `pulls' method that examined how changes in individual systematic effects changed the overall result.  

IceCube has also measured the atmospheric $\nu_e$ flux from 80 GeV to 6 TeV.  This analysis selected contained events and found the cascade flux was consistent with conventional expectations \cite{Aartsen:2012uu}.

\begin{figure}[!t]
  \centering
\includegraphics[width=3 in]{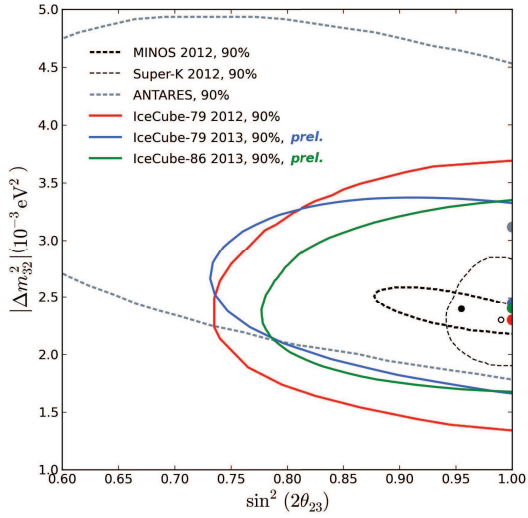}
  \caption{The allowed regions for different IceCube atmospheric neutrino oscillation analyses, shown along with limits from MINOS \cite{MINOS}, Super-K \cite{SuperK} and ANTARES \cite{AdrianMartinez:2012ph}.}
  \label{fig:oscillations}
 \end{figure}

At energies below 50 GeV, neutrino oscillations affect the $\nu$ flavor composition, primarily turning $\nu_\mu$ into $\nu_\tau$ which appear as cascades. For vertically upward-going neutrinos the first oscillation minimum occurs at 24 GeV.  Several IceCube analyses are studying $\nu_\mu$ disappearance, using contained events observed in DeepCore  \cite{Aartsen:2013jza}.  Figure \ref{fig:oscillations} shows the allowed regions in $\sin^2(2\theta_{23})$ and $\Delta m^2_{32}$ for three studies, along with the results from other experiments.   The three analyses use different techniques to select and reconstruct events, and determine the neutrino energy and zenith angle, with different tradeoffs between accuracy, efficiency and systematic uncertainties.

Looking ahead, an expanded collaboration is developing a proposal for a higher density infill array - Precision IceCube Next Generation Upgrade (PINGU), which will reduce our neutrino energy threshold to a few GeV \cite{IceCube:2013aaa}. PINGU will consist of 20-40 new strings, with a roughly 20 m spacing.  Each string will support 60-100 optical modules, spaced a few meters apart, instrumenting an effective mass of about 10 megatons.  The optical modules will be similar to those used in IceCube. 

The physics focus for PINGU is to use atmospheric neutrinos to determine the neutrino mass hierarchy by determining which neutrino species is heaviest.   It is sensitive to the mass hierarchy is through the Mikheyev\textendash Smirnov \textendash Wolfenstein (MSW) effect, whereby $\nu_1$ (the mass eigenstate that is largely $\nu_e$) traversing the Earth can resonantly interact with the electrons and be converted to another flavor.  The details of this conversion depend on whether the mass hierarchy is `normal' (with $\nu_1$ the lightest) or `inverted' (with $\nu_1$ the heaviest).   A definitive measurement of the mass hierarchy will require stringent control of the systematic uncertainties.

\section{Extra-terrestrial Neutrinos}

Many methods have been proposed to detect extra-terrestrial neutrinos.  These include point source searches, both continuous and episodic and multiple approaches to search for diffuse neutrinos.  IceCube is pursuing most of these methods. 

\begin{figure}[!t]
  \centering
\includegraphics[width=3.4 in]{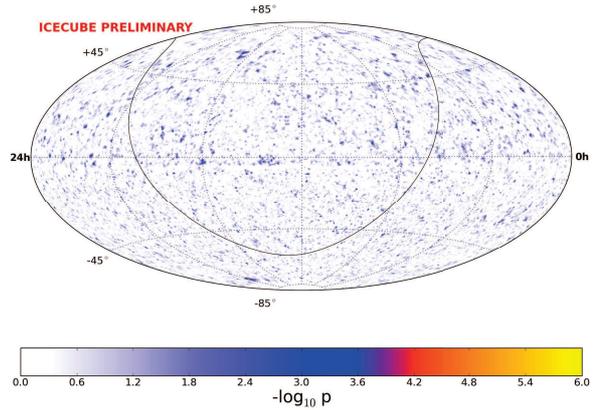}
 \caption{A neutrino sky map containing 390,000 events, collected over 1371 live days, with the full and partially completed IceCube detectors. There are no statistically significant excesses.  In the Northern Hemisphere, about 90\% of the events are from neutrinos, while in the Southern Hemisphere, they are mostly cosmic-ray muons.}
 \label{fig:pointsource}
\end{figure}

Figure \ref{fig:pointsource} shows a neutrino sky map with 390,000 events, collected over four years of detector operation \cite{pointsource}. The distribution is consistent with background expectations without any statistically significant excesses.  In the North, about 90\% of the events are atmospheric neutrinos, while inthe South, the triggers are mostly muons.  We perform an unbinned maximum-likelihood search and set declination-dependent upper limits to the flux from possible point sources.  In the most sensitive region (near the horizon), these limits reach the level of $E^2\phi < 10^{-12}$ TeV cm$^{-2}$ s$^{-1}$ for a source with an assumed $E^{-2}$ spectral index \cite{Aartsen:2013uuv}.   We have also performed dedicated searches for neutrinos from cosmic-ray production in blazars, the Cygnus region (identified by MILAGRO as a source of TeV gamma rays) and open star clusters \cite{stacking}, and from flares of active galactic nuclei \cite{flares}. 

We have also performed a variety of searches for episodic sources, including searches for gamma-ray bursts (GRBs) and for periodic or flaring sources.  The GRB study searches for events in temporal and spatial coincidence with GRBs observed by different satellites and reported by the Gamma-ray Coordination Network (GCN).  A search using 4 years of data found no significant correlation, and set limits a factor of two tighter than our previous search \cite{GRB}.  Recent theoretical calculations predict lower neutrino fluxes; the new experimental limits are comparable to the newer calculations. 

IceCube is also pursuing multimessenger astronomy, correlating neutrino arrival times and directions with data from optical and gamma-ray telescopes.  We also send out alerts to these telescopes when neutrino pairs (within a selected time/space window) are observed \cite{multimessenger}.  Currently, we have arrangements with ROTSE, the PTF survey at the Palomar Observatory, MAGIC and VERITAS.  To date, no significant correlations have been observed.  A future enhancement will send triggers when very high energy $\nu_\mu$ or $\nu_\tau$ events are observed.  

Searches for diffuse extra-terrestrial neutrinos depend on finding a clear excess over the expected atmospheric background.  There are several  handles for differentiating atmospheric and extra-terrestrial neutrinos.  For example, they are expected to have a harder energy spectrum.  Fermi shock acceleration predicts an $E^{-2}$ spectrum, compared to $E^{-3.7}$ \& $E^{-4.0}$ for conventional atmospheric neutrinos above and below the knee of the spectrum (about 400 TeV for neutrinos), and  $E^{-2.7}$ \& $E^{-3.0}$ for prompt neutrinos.  The flavor composition is also different, with extra-terrestrial neutrinos expected to be $\nu_e$:$\nu_\mu$:$\nu_\tau$ = 1:1:1, while atmospheric neutrinos are mostly $\nu_\mu$.  Finally, energetic downward-going atmospheric neutrinos are expected to be accompanied by an air shower, including energetic muons.

\begin{figure}[!t]
\centering
\includegraphics[width=3 in]{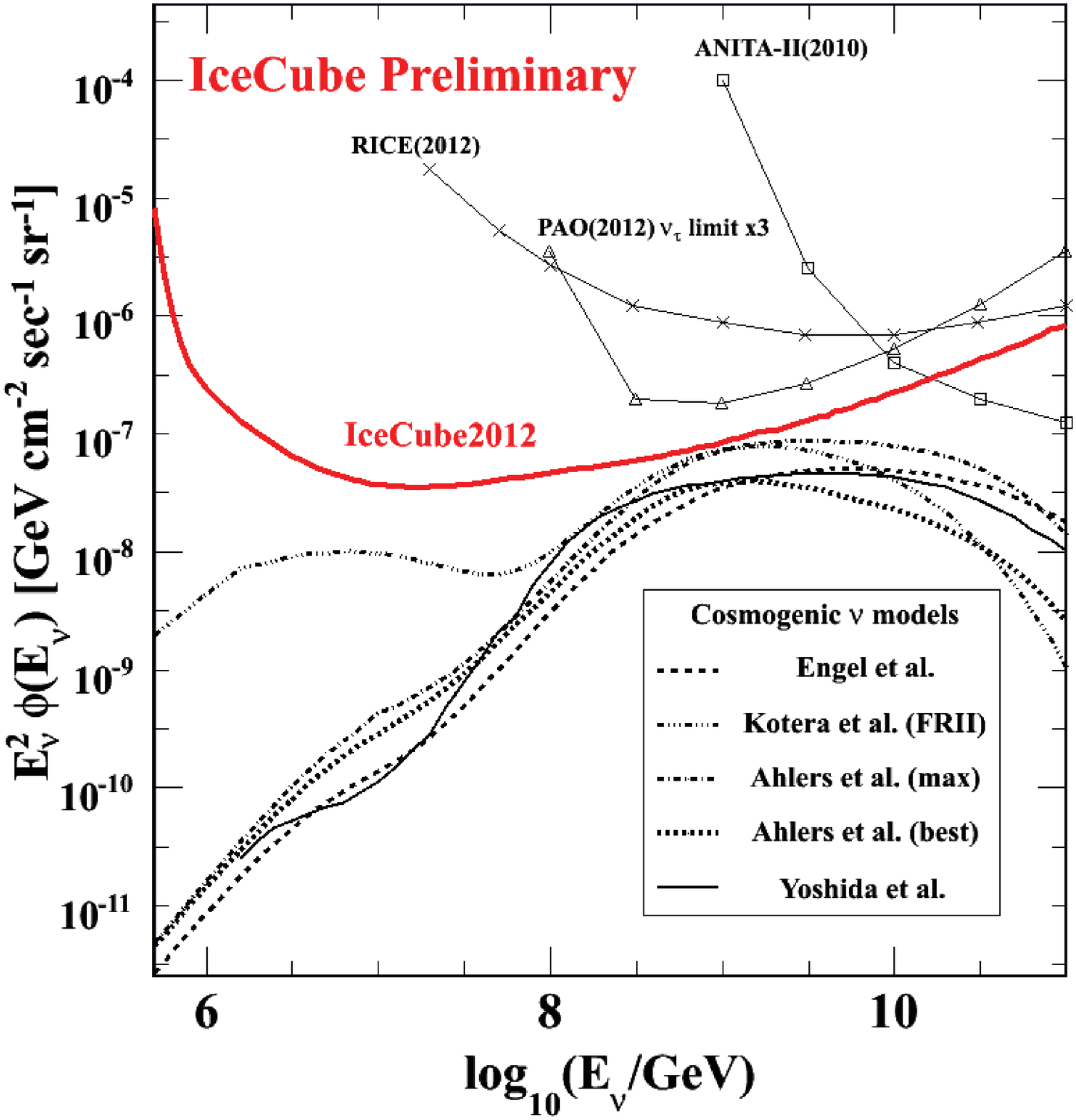}
 \caption{Differential limits on the flux of GZK neutrinos, set with two years of IceCube data, and compared with limits from RICE \cite{RICE}, the Pierre Auger Observatory \cite{PAO} and ANITA \cite{ANITA}. Also shown are several theoretical calculations.}
\label{fig:GZK}
\end{figure}

A recent search for extremely high energy neutrinos in IC79 and IC86 found Bert and Ernie, two neutrino-induced cascades (showers) with energies above 1 PeV \cite{Aartsen:2013bka}.  These events are incompatible with an atmospheric origin at the 2.8$\sigma$ level.  They are also too low in energy to be plausible Greisen\textendash Zatsepin\textendash Kuzmin (GZK) neutrinos from the interaction of ultra-high energy (above about $4\times 10^{19}$ eV) protons with cosmic microwave background radiation photons.  We set the differential limits shown in Fig. \ref{fig:GZK} on the flux of ultra-high energy neutrinos, following an assumed GZK spectrum.  The differential limit includes Bert and Ernie, which cause the decrease in the limits around 1 PeV.  

We have also searched for extra-terrestrial $\nu_\mu$ using data from IC59 \cite{numu}. This search looked for a hard component to the $\nu_\mu$ energy spectrum visible above the softer atmospheric neutrino flux.  Finding none, a 90\% confidence level flux limit was set for an $E^{-2}$ spectrum at $E^2\phi < 1.4\times 10^{-8}$ GeV/cm$^2$/s/sr.    A 90\% confidence level  limit was also set at  on the prompt flux, at 3.8 times the central value of the prediction in Ref. \cite{Enberg:2008te} ("ERS").  

 Another search of the same data set selected events that were contained within the detector and had a topology compatible with a electromagnetic or hadronic shower.  The energy spectrum of selected events was fit to a mixture of atmospheric muons and neutrinos (with separate components for prompt and conventional $\nu$) and an extra-terrestrial component. The data was compatible with the combined atmospheric expectation, and a 90\% confidence level flux limit was set for an $E^{-2}$ spectrum, at $E^2\phi < 1.7\times 10^{-8}$ GeV/cm$^2$/s/sr \cite{nue}. A limit was also set on the prompt flux, at 9 times the ERS central value.  Similar searches are now underway using IC79 and IC86. 

\begin{figure}[!t]
\centering
\includegraphics[width=3 in]{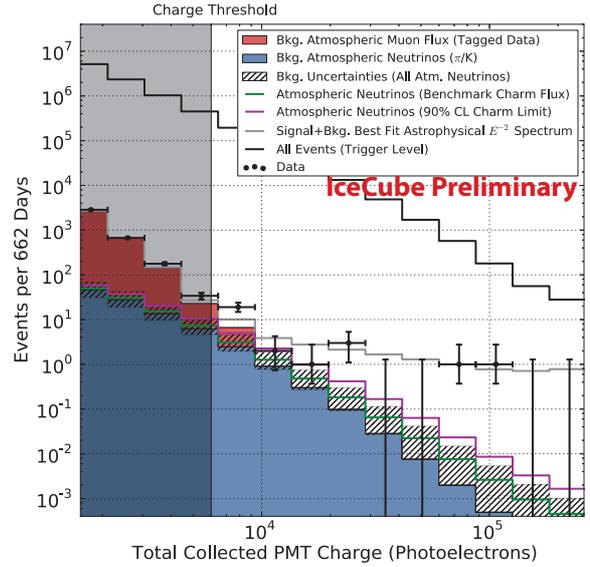}
 \caption{Total event charge in photoelectrons (a proxy for energy) for the 28 contained events.  The points with error bars are the data.  The blue shows the atmospheric neutrino background, with the hatched area showing the uncertainty due to the prompt neutrino flux.  The red curve shows the atmospheric muon background, while the grey curve shows the background plus an astrophysical signal. Only events with more than 6,000 photoelectrons were used in the quantitative analysis.}
\label{fig:HESEcharge}
\end{figure}

A follow-on to the EHE analysis discussed above searched for additional neutrinos with energies around 1 PeV in the same two years of data.  It incorporated several new features, including a method for determining the atmospheric muon background directly from the data and a calculation of the probability that a down-going atmospheric neutrino will be accompanied by muons which will cause the event to be rejected as a neutrino candidate \cite{Claudio}.  This search divided the detector into a central active region and an outside veto region and required that the event originate in the 400 megaton active region.  It selected events containing more than 6,000 observed photoelectrons, without differentiating between event topologies.  The search found 28 events in two years of data, including the previously identified Bert and Ernie.  Figure \ref {fig:HESEcharge} shows the observed charge distribution, in photoelectrons, for the selected events.

The background from cosmic-ray muons was estimated from the data by using an expanded double-layer veto.  By counting the number of events tagged in one veto layer but not the other, the cosmic-ray muon background was estimated to be $6.0\pm 3.3$ events.  The atmospheric neutrino background was estimated based on IceCube measurements at lower energies, where any extra-terrestrial contamination was expected to be minimal \cite{numu}.  This number was scaled by the probability that the atmospheric neutrino would not be accompanied by an atmospheric muon or bundle.   This probability was estimated from a Monte Carlo calculation, with a 'floor' added, requiring that there was at least a 10\% probability of being unaccompanied.   With this, the estimated background from conventional atmospheric neutrinos was $10.6^{+5.0}_{-3.6}$ events.  Since this was extrapolated from lower energy measurements, it does not include prompt neutrinos.  The prompt flux was estimated to be 1.5 events, based on the central ERS value \cite{Enberg:2008te}, with an adjustment to include an improved, composition-dependent cosmic-ray knee.    Because of the large uncertainties in the prompt flux, we also considered a background model with the prompt flux at the 90\% confidence level limit given above. We also considered a-priori and a-posteriori significance calculations, using respectively the 26 events first observed in this analysis ({\it i.e.} excluding Bert and Ernie, which prompted the study), and on all 28 events.  From these studies, we have observed evidence for extra-terrestrial neutrinos at the $4\sigma$ level.

\begin{figure}[!t]
\centering
\includegraphics[width=3 in]{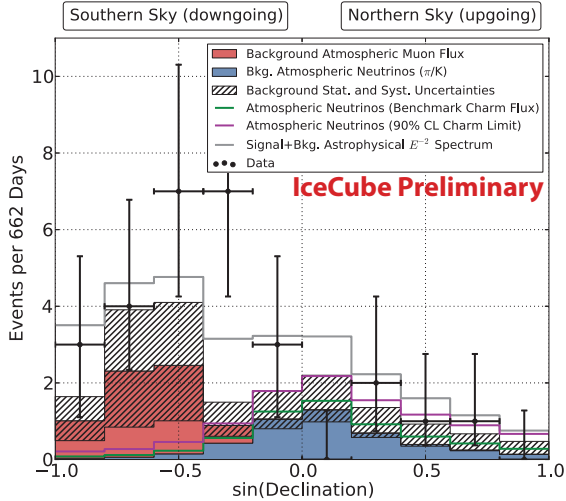}
\caption{Zenith angle distribution for the 28 contained events.  The points with error bars are the data.  The red region shows the atmospheric muon background; these are mostly down-going.  The blue shows the mostly-upgoing atmospheric neutrino background, with the hatched region again showing the uncertainty due to the prompt neutrino flux, while the grey includes an astrophysical flux.  The signal excess is almost entirely down-going, inconsistent with an excess atmospheric neutrino flux but consistent with an extra-terrestrial signal.
 \label{fig:HESEzenith}}
\end{figure}

Figure \ref{fig:HESEzenith} shows the zenith angle distribution of the 28 events, along with the expected signal and background distributions.  The atmospheric muon background is entirely downward-going.  Because many downward-going atmospheric neutrinos are vetoed, and energetic upward-going neutrinos may be absorbed in the Earth, the conventional atmospheric neutrino background is concentrated around the horizon.  The fitted extra-terrestrial signal is more downgoing than upgoing because of neutrino absorption in the Earth, but still less asymmetric than the data.  The distribution of 24 downward-going events and 4 upward-going ones is about 1.5$\sigma$ away from expectations for an isotropic excess, as expected from a diffuse astrophysical signal. 

\begin{figure}[!t]
\centering
\includegraphics[width=3 in]{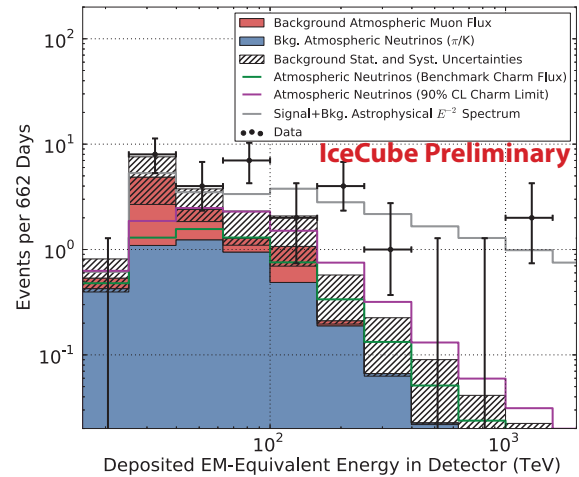}
\caption{Visible energy deposition for the 28 contained events assuming that all of the energy is electromagnetic.  Hadronic showers produce 10-15\% less light, depending on energy.  The points with error bars are the data.  The red shows the atmospheric muon background, which is concentrated at energies below 60 TeV.   The blue shows the mostly-upgoing atmospheric neutrino background, with the hatched region again showing the uncertainty due to the prompt neutrino flux, while the grey includes an astrophysical flux.  
 \label{fig:HESEenergy}}
\end{figure}

Figure \ref{fig:HESEenergy} shows the visible energy distribution of the 28 events, assuming that the deposited energy is electromagnetic in origin.  Most of the muon background is at energies below 60 TeV, since the muon energy spectrum is very soft, and muons are increasingly unlikely to pass the veto selection with increasing energy.   The conventional atmospheric neutrinos have a soft spectrum ($E^{-3.7/4.0}$), while the prompt flux is harder ($E^{-2.7/3.0}$), but still softer than the expected $E^{-2}$ astrophysical flux.   We have fitted the data at energies above 60 TeV, where the atmospheric muon background should be negligible.  At energies up to 1 PeV the excess over the atmospheric neutrino flux is roughly compatible with an $E^{-2}$ spectrum, at a level about $1.2\pm0.4 \times10^{-8}$  GeV/cm$^2$/sr/s.  The absence of events at energies much above 1 PeV requires that either the $E^{-2}$ spectrum is cut off at an energy below 10 PeV, or that the spectrum is softer than $E^{-2.0}$. An $E^{-2.2}$ spectrum would fit the data adequately.  

We have studied the arrival directions of the 28 events and find no statistically significant clustering.  We also observe no statistically significant association with the galactic plane. 

\begin{figure}[!t]
\centering
\includegraphics[width=3 in]{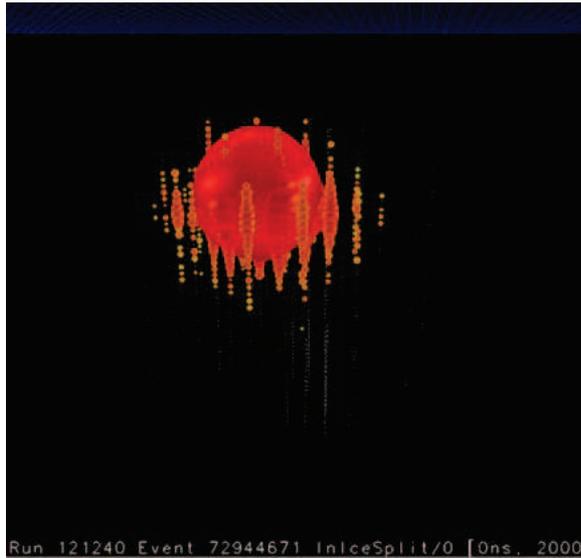}
\caption{Event display showing Big Bird, with 378 optical modules hit. Each sphere shows a hit optical module.  The size of the spheres shows the number of photoelectrons observed by the DOM, while the color indicates the time, with red being earliest, and blue latest.
\label{fig:BigBird}}
\end{figure}

Looking ahead, we are starting to examine the data collected in 2012.  One very high energy event, Big Bird, appeared in the 10\% of the data that we use to tune our cuts.  It is shown in Fig. \ref{fig:BigBird}.  A total of 378 DOMs were hit, making it the brightest neutrino event seen yet.

\section{Conclusions}

Since its completion in December, 2010 IceCube has been running smoothly, with 98.5\% of the deployed DOMs collecting data, and a typical up-time of 99\%.   Using data gathered before and after detector completion, IceCube is studying a wide variety of topics involving cosmic-rays, searches for weakly interacting massive particles, exotic physics, and atmospheric and extra-terrestrial neutrinos.  We have made a precise measurement of the cosmic-ray energy spectrum at energies between 1.58 PeV and 1.26 EeV, and measured the cosmic-ray anisotropy and composition.  We have searched for neutrinos from dark matter annihilation in the Sun, galactic center and halo, and dwarf galaxies, and set limits on magnetic monopoles.

We have made detailed measurements of the atmospheric $\nu_e$ spectrum up to 6 TeV, and the $\nu_\mu$ flux up to 1 PeV.  We observe an excess over the expected atmospheric neutrino background, with a significance at the $4\sigma$ level.  The excess is compatible with an $E^{-2}$ extra-terrestrial flux at a level about $1.2\pm0.4 \times10^{-8}$  GeV/cm$^2$/sr/s, with a cutoff energy above 1 PeV; alternately, it is compatible with a somewhat softer spectrum.   

We have observed atmospheric neutrino oscillations.  Looking ahead, the PINGU subarray will extend IceCube's threshold down to a few GeV, allowing for the observation of resonance conversion of electron neutrinos (the MSW effect), and, systematic uncertainties permitting, a determination of the neutrino mass hierarchy. 

\vspace*{0.5cm}
\footnotesize{\bf Acknowledgments:}{\ We thank the organizers for putting together a very enjoyable conference. This work was supported in part by U.S. National Science Foundation under grant 0653266 and the U.S. Department of Energy under contract number DE-AC-76SF00098.}

\end{document}